\documentstyle[aps,prl,psfig,epsf]{revtex}
\newcommand{\kdp}{\mbox{KH$_2$PO$_4$} }
\begin{document}

\twocolumn[\hsize\textwidth\columnwidth\hsize\csname@twocolumnfalse\endcsname

\title{Ferroelectricity and isotope effects in hydrogen-bonded KDP crystals}

\author{S. Koval$^{\rm (1,2)}$, J. Kohanoff,$^{\rm (3)}$,
        R. L. Migoni $^{\rm (2,1)}$, and E. Tosatti$^{\rm (1,4)}$ }

\address{
         $^{1)}$ International Centre for Theoretical Physics,      \\
                 Strada Costiera 11, I-34014 Trieste, Italy         \\
         $^{2)}$ Instituto de F\'{\i}sica Rosario, Universidad
                 Nacional de Rosario,                               \\
                 27 de Febrero 210 Bis, 2000 Rosario, Argentina.    \\
         $^{3)}$ Atomistic Simulation Group, The Queen's University,
                 Belfast BT7 1NN, Northern Ireland.\\
         $^{4)}$ International School for Advanced Studies (SISSA),      \\
                 and Istituto Nazionale Fisica della Materia (INFM),
                 Via Beirut 4, I-34014 Trieste, Italy
                 }
\date{\today}
\maketitle

\begin{abstract}

Based on an accurate first principles description of the energetics in H-bonded
KDP, we conduct a first study of nuclear quantum effects and of the changes 
brought about by deuteration. Cluster tunneling involving also heavy ions is 
allowed, the main effect of deuteration being a depletion of the proton 
probability density at the O-H-O bridge center, which in turn weakens its 
proton-mediated covalent bonding. The ensuing lattice expansion couples 
selfconsistently with the proton off-centering, thus explaining both the giant 
isotope effect, and its close connection with geometrical effects.

\end{abstract}

\vspace{0.5cm}
]

Potassium dihydrogen phosphate (KH$_2$PO$_4$, or KDP) belongs to a family of
ferroelectric (FE) crystals where molecular units are linked by hydrogen bonds,
the ferroelectricity being connected to proton off-center ordering in the bonds.
A characteristic feature of this family is the large increase in the Curie
temperature T$_c$ upon deuteration. In this particular case, it goes from
$\simeq$ 122 $K$ in KDP to $\simeq$ 229 $K$ in the deuterated compound (DKDP).
The origin of this huge isotope effect is still controversial, and has been
mostly understood in terms of the quantum tunneling model proposed by
Blinc,\cite{Bli60} later modified by inclusion of the coupling between proton
motion and the K-PO$_4$ dynamics.\cite{models} While direct experimental
indications of tunneling have recently emerged\cite{reiter}, the connection
between proton tunneling and isotope effect remains unclear. There is in fact 
strong evidence,\cite{geom_evid,McM90n} that isotope substitution acts rather
through a {\em geometrical} modification of the hydrogen bonds,~\cite{Ubb39}
with an expansion of the O-H-O distance. The proton off-centering, and thus the
corresponding increase of lattice parameter upon deuteration, appear to
be remarkably correlated to the increase of order parameter and of T$_c$.
These findings stimulated new theoretical work where some of these facts could 
be addressed without invoking tunneling,\cite{Sug96} however so far only at a 
rather phenomenological level.

In the first part of this letter we investigate, using electronic structure 
calculations within Density Functional Theory (DFT), the relationship between 
proton ordering, polarization, and geometry in KDP. In the second part, we
introduce a study of energy and subsequently the quantization of the collective
ion displacements in small KDP clusters, embedded in a host paraelectric 
lattice. These calculations demonstrate the difference between deuterated and 
protonated KDP, the more delocalized proton bridging the oxygens and pulling 
them together more effectively than the deuteron. This phenomenon, which is 
at the root of the geometric effect, is further illustrated by solving in the 
last part a selfconsistent nonlinear model.

For the DFT calculations we use two different approaches: one employing a basis
set of confined pseudoatomic orbitals (SIESTA),\cite{Ord96} another a plane 
wave (PPW) representation.\cite{Cav99} For the first we choose a double-zeta 
basis set with polarization functions, and an orbital confinement energy 
$E_c=50$ meV. In the second, we set the energy cutoff to 150 Ry. In both cases
exchange-correlation terms are computed using a gradient-corrected (GGA)
functional,\cite{Per96} and norm-conserving pseudopotentials~\cite{Tro91} are
employed to represent the interaction between ionic cores and valence
electrons. We also include the nonlinear core corrections for a proper 
description of the K ion. The Brillouin zone $\Gamma$-point alone provides a 
sufficient sampling in the large supercells used.

The paraelectric (PE) structure of KDP is body-centered tetragonal with two
KH$_2$PO$_4$ units per lattice 

\vspace{-0.3 truecm}
\begin{figure}[b]
\epsfxsize=7.5cm
\epsfysize=4.8cm
\hspace*{0.cm} \epsfbox{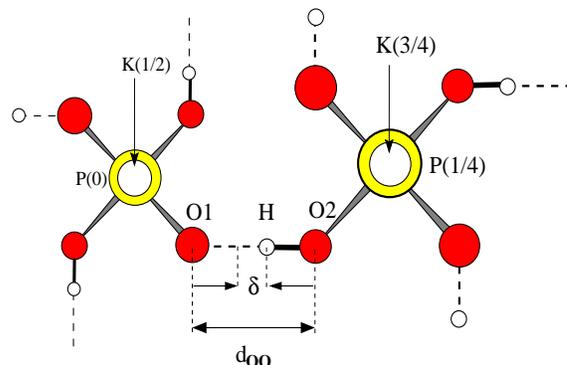}
\vskip 0.cm
\caption{Schematic view of the internal structure of KDP along the tetragonal
axis. P and K coordinates relative to $c$ along this axis are indicated.
Covalent and H-bonded hydrogens in the FE phase are attached to the
corresponding oxygen through full and broken lines, respectively.}
\end{figure}

\noindent 
site. We use the conventional $bct$ cell doubled along the 
tetragonal $c$ axis (64 atoms) to describe homogeneous
distortions, and a conventional $fct$ cell doubled along the $c$-axis (128 
atoms) for local distortions. The internal structure is depicted in Fig. 1. 
Above T$_c$ the protons occupy with equal probability two equivalent off-center
positions in the H-bond separated by a distance $\delta$,\cite{Nel87} while 
below T$_c$ they order in such a way that each PO$_4$ group has two covalently 
bonded and two H-bonded hydrogen atoms.

We first analyse the relationship between proton ordering and polarization. 
We start from the average experimental structure of the PE phase of KDP at 
T$_c^{KDP}$+5~K,\cite{Nel87} with the hydrogens centered in the O-H-O bonds.
By fully relaxing the atomic positions in the $bct$ cell, the H atoms move 
collectively off-center towards the O2 oxygens, and away from O1, as indicated 
in Fig. 1. Analysis of Mulliken populations and charge density differential 
maps~\cite{koval} indicates that the off-centering of the H atoms induces a 
covalent charge displacement from O2 towards the O2-H bonds, whereas the O1-H 
bonds weakens into a hydrogen bond. In addition, there is a charge flow from 
the O2-P to the O1-P bonds, accompanied by an increase of the P-O2 and a 
decrease of the P-O1 distances.  P atoms are thus driven off-center in the 
PO$_4$ tetrahedra, which polarize further. Unbalanced electrostatic forces 
induce a displacement of the K$^+$ ions along the $c$-axis, towards the 
charge-excess (O1) side of PO$_4$ units. A detailed description of the 
classical FE distortion in KDP can be found in Ref.~\cite{koval}. These results
are also in agreement with Ref.~\cite{Zha02}. In order to identify the driving 
mechanism of the FE instability, we investigate the {\it ab initio} potential 
energy surface (PES) as a function of the proton off-centering parameter 
$\delta=d_{OO}-2d_{OH}$, and of the K-P relative displacement along $c$, 
$\gamma=c-2d_{KP}$, which provides a measure of the polarization.~\cite{unpub}
By fully relaxing the oxygen positions for each chosen $(\delta, \gamma)$ pair,
we obtain a two-dimensional double-well PES with a saddle point at $\delta = 
\gamma = 0$, whose contours are reported in the inset to Fig. 2. According to 
this PES, the crystal is stable against polarization ($\gamma \neq 0$) unless 
the protons are ordered off-center ($\delta \neq 0$). In Fig. 2 we show cuts of
the PES at different values of $\gamma$, indicating that, even for vanishing 
$\gamma$, the energy minimum corresponds to a finite $\delta$, i.e. protons are
always collectively unstable at the H-bond centers. Therefore, we confirm that 
the source of the FE instability is indeed the H off-centering.

In order to address next the quantum effects, in particular the isotope
substitution, we need to introduce quantum fluctuations in the above classical
picture. Barring for the time being a full brute force quantum mechanical 
calculation for all the ionic degrees of freedom of the infinite system, we 
take a different approach to the problem. Important quantum effects are 
identified as those involving correlated H motions (as shown in Fig. 1) with 
relaxation of K and P ions, most favorable for exhibiting 

\begin{figure}[tb]
\vspace*{-1.7 truecm}
\hspace*{-0.9 truecm}
\psfig{figure={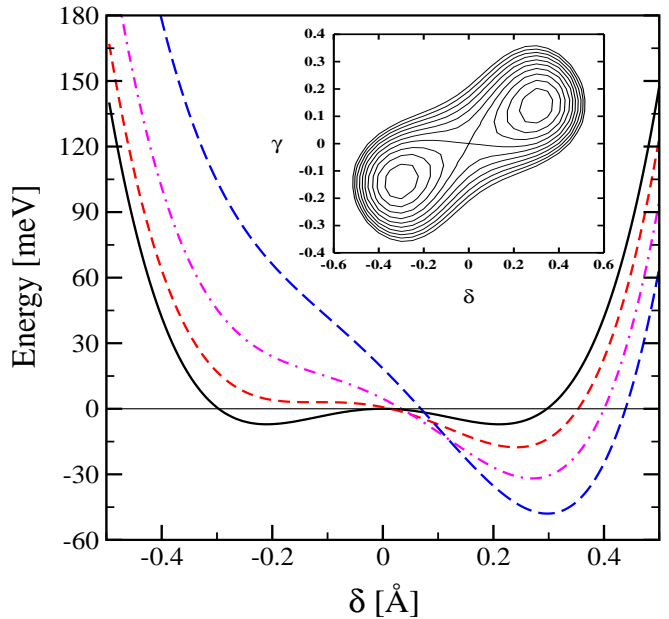},angle=0,width=10.5cm,height=12.9cm}
\vspace{-3.5truecm}
\caption{Energy profiles as a function of $\delta$, for values of
$\gamma=$ 0 (solid line), 0.02 (short-dashed), 0.05 (dot-dashed) and
0.1 (long-dashed) {\AA}. The minima are always at
$\delta\neq 0$, and for $\gamma\geq 0.02$ {\AA} the curves exhibit a single
minimum. The inset shows equispaced energy contours (step = 13.6 meV/
\kdp unit). The minima at $(\delta, \gamma) \simeq \pm (0.3, 0.15) \AA$
lie $\simeq 50$ meV below the saddle point at (0,0).}
\end{figure}

\noindent 
FE instabilities. We shall consider this correlated pattern as a single 
collective coordinate $ \delta_c$. Quantization of this coordinate will 
moreover be carried out within a finite-size cluster. Even though
quantum fluctuations span in principle all length scales, those that occur at 
short range should be sufficiently revealing at least away from critical 
points.

In this spirit, we consider a series of small KDP clusters of increasing size, 
embedded inside an undistorted host PE matrix. For these clusters, we first 
determine, classically, the total energy variation as a function of 
$\delta_c/2$. The clusters comprise N hydrogens, in the following order : 
(a) N=1; (b) N=4, fully connecting a PO$_4$ group to the host; (c) N=7,
connecting two PO$_4$ groups; (d) N=10, connecting three PO$_4$ groups. In 
order to ascertain the effect of the volume increase seen upon deuteration, we 
repeat the calculations by expanding the host structural parameters to the 
corresponding experimental values of DKDP at T$_c^{DKDP}$+5~K.\cite{Nel87} The 
results show (Fig. 3, solid curves) that there is an instability only when the 
cluster size is sufficiently large, thus providing a measure of the FE 
correlation length. The instability is much stronger (note the larger energy 
scale), and the correlation length accordingly shorter, with the expanded 
structural parameters of DKDP. This is in close agreement with the experimental
trend, showing that the FE order grows with volume. Finally, the involvement of
P and K motions is important. The dashed curves in Fig. 3 represent the energy 
of a distortion where only H atoms are allowed to move, all other atoms being 
kept fixed. The much higher energies indicate that, not surprisingly, this kind
of oversimplified distortion is far from real, leaving, e.g., the KDP lattice 
fully stable.

The next step in order to study the quantum effects due to proton or deuteron 
zero point motion, is to quantize the clusters with respect to the local 
collective coordinate $\delta_c/2$. The corresponding kinetic energy involves 
the mass of each ion proportionally to the square of its displacement, and will
change upon deuteration. We find that the deuterium (hydrogen) effective mass 
for this correlated motion is about $\mu_D = $ 3.0 ($\mu_H=$ 2.3) proton masses
in DKDP (KDP) clusters. Solving Schr\"odinger's equation for the single 
variable $\delta_c/2$ with mass $\mu_{H,D}$ and effective potentials of Fig. 3
is trivial. The ground state (GS) energy levels which lie below the energy 
barrier at $\delta_c/2 =0$  are indicated by dotted lines. A negative energy 
signals the occurrence of tunneling between + and - $\delta_c/2$, and its onset
provides a rough indication of the correlation volume. The results show that it
comprises more than N=10 hydrogens in KDP, but no more than N=4 deuteriums in 
DKDP. In both systems, we eventually expect the collective tunnel splitting to 
converge to zero as $N \rightarrow \infty$, signaling long-range 
ferroelectricity; the tunnel splitting should conversely remain nonzero in the 
quantum paraelectric state, attainable at high pressure.\cite{McM90n,Nel91}

Remarkably, deuteration at fixed geometry does not affect the fact that the 
tunnel splittings in large clusters~\cite{Sam73} are exceedingly smaller than 
k$_B$T$_c$, which renders T$_c$ practically insensitive. In addition, the 
effective mass change is reduced by the involvement of the heavy nuclei. This 
agrees with high-pressure experiments,\cite{McM90n,Nel91} where under fixed 
structural conditions the isotope effect appears to be only very slight.

The fact that energy barriers in DKDP are much larger than those in KDP implies
that quantum effects are significantly reduced in the expanded DKDP lattice. 
After this observation, the reason for the geometric effect becomes clear: in 
all KDP clusters considered here the wavefunction (WF) for the collective 
coordinate $\delta_c$ is much more delocalized, with a  larger amplitude near 
the H-centred position $\delta_c=0$ between O1 and O2, than in DKDP. In KDP in 
other words, zero-point motion pushes protons towards the center, much more 
effectively so than deuterons in DKDP. An increased probability for the proton 
to bridge midway between the two oxygens is not irrelevant to the crystal
cohesion, and we identify precisely that as the element which compresses the
cell from a larger classical value to the smaller value found for KDP. To 
estimate an upper limit to that effect, we compare the lattice parameter and 
the O-H-O bridge length $d_{OO}$ of two classical electronic calculations: one 
with the hydrogens forced to sit in the bridge center, the other with H fully 
off-center in the classical FE state. The result is $d_{OO}$ = 4.51 {\AA} when 
H is off-center, dropping to $d_{OO}$ = 

\begin{figure}
\vspace*{-1.6 truecm}
\centerline{
\psfig{figure={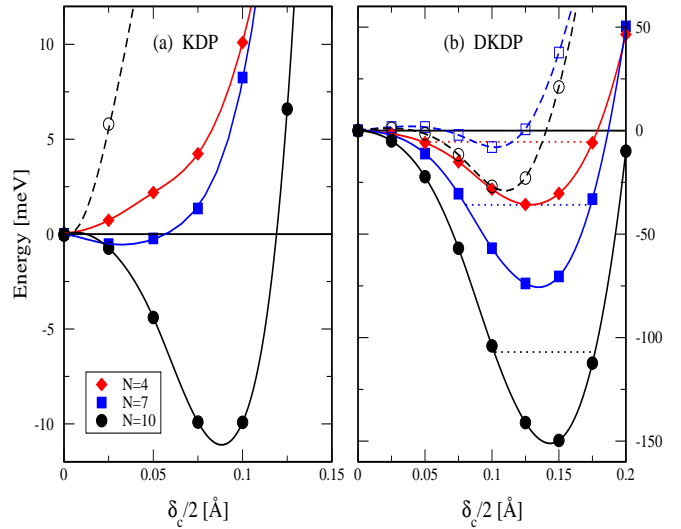},angle=270,width=9.cm,height=8.3cm}}
\vspace{-0.25truecm}
\caption{{\it Ab initio} energy profiles for classical distortions of
clusters embedded in undistorted PE structures of (a) KDP
and (b) DKDP. Reported are clusters of: N=4 (diamonds), N=7
(squares), and N=10 (circles). Full symbols and solid
lines: P and K positions inside the cluster are allowed to relax as H are
off-centered; empty symbols and dashed lines: only H atoms are moved.
Dotted lines: GS energies (only negative ones, signaling tunneling).}
\end{figure}

\noindent  
4.42 {\AA} when H is centered. At the equilibrium volume, we estimate that
proton centering creates an equivalent pressure of approximately 20 kbar. Thus,
a centered hydrogen acts as a very strong attraction center that pulls the two 
bridge oxygens together, the whole lattice volume shrinking by 2.3 \% as a
result. In the true high-temperature PE phase the hydrogens are of course not
exactly centered, but appear on both sides of the H-bond, thus reducing the 
magnitude of the effect.

Deuteration of KDP to DKDP at fixed geometry results in a depleted probability
for the deuterium to reach the bridge center, as shown in Fig 4(a) where we 
plot the proton and deuteron WF in the DKDP potential of the N=7 cluster of 
Fig. 3(b). This depletion releases a fraction of the O-H-O bond grip, causing a
small lattice expansion which has the effect of increasing the potential wells 
depth, making the deuteron even more localized, and so on in a self-consistent 
manner. The overall self-consistent effect is eventually much larger than the
mere fixed-geometry replacement of the proton with the deuteron mass.
In fact, for the same cluster size in KDP, instead of a double-peak, the WF
exhibits a broad single peak centered in $\delta_c/2 = 0$ (see Fig. 4(a)). This
mechanism is now capable of explaining, at least qualitatively, the increase in
the order parameter and T$_c$, which are related to the mass only indirectly,
through the geometric modification of the length and energy scales.

We constructed a simple model that demonstrates the nonlinear behaviour arising
from isotope substitution in KDP by adding to the effective hydrogen potential
in the 

\begin{figure}
\centerline{
\psfig{figure={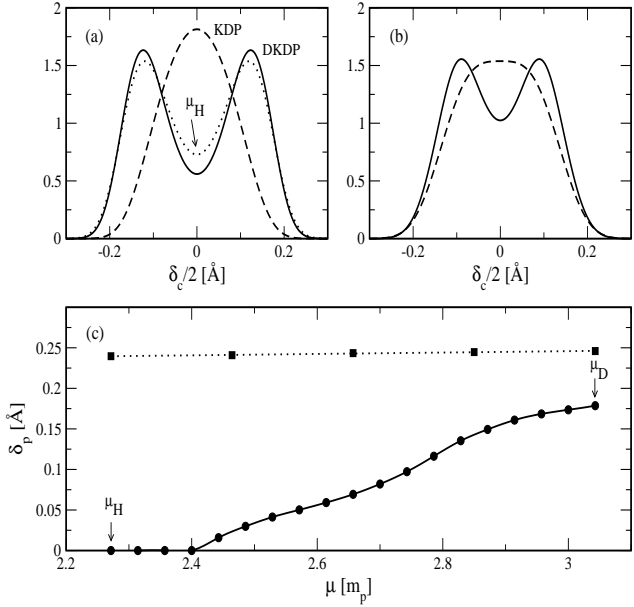},angle=270,width=9.cm,height=8.3cm}}
\caption{WF in the N=7 cluster PES for (a) {\it ab initio} and (b)
self-consistent model calculations. Solid (dashed) lines are for DKDP (KDP).
Dotted line is for H in the DKDP PES. (c) WF peak separation
$\delta_p$ as a function of the effective mass $\mu$ (given in units of the
proton mass) for the self-consistent
model (circles) and for fixed DKDP potential (squares).
Lines are guides to the eye.}
\end{figure}

\noindent 
cluster Schr\"odinger equation a quadratic WF-dependent 
term of the form $V_{\rm eff}(x)=V_0(x) - k |\Psi(x)|^2$, where $x=\delta_c/2$
and $V_0(x)$ is a quartic double-well similar to those of Fig. 3. The nonlinear
$|\Psi(x)|^2$ term signifies the positive feedback discussed above, since a 
decreasing mass will increase $|\Psi(x)|^2$ at the center, which will in turn
lower the barrier at that point, further increasing $|\Psi(x)|^2$, and so on.
By choosing appropriately $k$ and the ``bare'' potential $V_0(x)$, we find that
the corresponding {\it ab initio} profiles for KDP and DKDP can be 
qualitatively reproduced (Fig. 4 (b)). The self-consistent solution evolves 
from a double-peak to a single-peak situation when changing the mass from 
``pure DKDP'' to ``pure KDP'', as shown by the circles in Fig. 4 (c). Such a 
large mass dependence, in striking contrast with the very weak dependence
obtained at fixed DKDP potential and geometry (squares), can now explain the
large and controversial isotope effect in the ferroelectricity of KDP.

Summarizing, we confirm that the H off-center ordering is the source of FE 
instability. First principles calculations in a model PE phase show that 
distortions involving only H atoms do not display significant instabilities.
Instead, heavy cluster tunneling involving also K and P ions, explains the 
``double occupancy'' in the PE phase, in agreement with the observation of 
P-ion double-site distribution.~\cite{McM90e} Although tunneling is of course
the crucial ingredient in the isotope effect, its mere change at fixed
lattice geometry accounts only for a very small fraction of the full isotope 
effect. The main effect of replacing deuterons with protons appears to be an 
enhancement of the quantum delocalization of the proton in the bond center 
region. That gives rise to a strong proton-mediated covalency in the O-H-O 
bridge, which shrinks the lattice, which further delocalizes the proton, and 
so on in a nonlinear loop. In the end, the huge isotope effect on T$_c$ is 
dominated by the geometrical effect. This effect, however, is triggered by
tunneling, thus reconciling these two aspects which were largely debated in 
the past.

We thank R.J. Nelmes, M.I. McMahon, R. Resta, A. Bussmann-Holder, G. Colizzi,
G. Reiter, M.G. Stachiotti and D. Marx for helpful discussions. R.M. and S.K.
thank support from CONICET, Argentina, and from ICTP, Trieste. E.T.'s work
was also supported by MIUR COFIN01, and by INFM/G.

\end{document}